\documentclass[twocolumn]{aastex631}
\usepackage{amsmath}
\usepackage{CJK}
\newcommand\todo[1]{#1}
\shorttitle{Turbulent vortex probed by scattering-induced polarization in IRS 48}
\shortauthors{Yang et al.}
\graphicspath{{./}{figures/}}

\begin{document}
\begin{CJK*}{UTF8}{gbsn}

\title{Turbulent vortex with moderate dust settling probed by scattering-induced polarization in the IRS 48 system}

\correspondingauthor{Haifeng Yang}
\email{hfyang@pku.edu.cn}

\author[0000-0002-8537-6669]{Haifeng Yang (杨海峰)}
\altaffiliation{Boya Fellow}
\affil{Kavli Institute for Astronomy and Astrophysics, Peking University, Yi He Yuan Lu 5, Haidian Qu, Beijing 100871, People's Republic of China}

\author[0000-0001-5811-0454]{Manuel Fern\'andez-L\'opez}
\affiliation{Instituto Argentino de Radioastronom\'ia (CCT-La Plata, CONICET; CICPBA), C.C. No. 5, 1894, Villa Elisa, Buenos Aires, Argentina}

\author[0000-0002-7402-6487]{Zhi-Yun Li}
\affiliation{Department of Astronomy, University of Virginia, Charlottesville, VA 22903, USA}

\author{Ian W. Stephens}
\affiliation{Department of Earth, Environment, and Physics, Worcester State University, Worcester, MA 01602, USA}

\author[0000-0002-4540-6587]{Leslie W. Looney}
\affiliation{Department of Astronomy, University of Illinois, 1002 West Green Street, Urbana, IL 61801, USA}
\affiliation{National Radio Astronomy Observatory, 520 Edgemont Rd., Charlottesville, VA 22903, USA}

\author[0000-0001-7233-4171]{Zhe-Yu Daniel Lin}
\affiliation{Department of Astronomy, University of Virginia, Charlottesville, VA 22903, USA}

\author{Rachel Harrison}
\affiliation{Department of Astronomy, University of Illinois, 1002 West Green Street, Urbana, IL 61801, USA}
\affiliation{School of Physics and Astronomy, Monash University, Clayton VIC 3800, Australia}

\begin{abstract}
We investigate the crescent-shaped dust trap in the transition disk, Oph IRS 48, using well-resolved (sub)millimeter polarimetric observations at ALMA Band 7 (870 $\mu$m). 
The dust polarization map reveals patterns consistent with dust scattering-induced polarization. There is a relative displacement between the polarized flux and the total flux, which holds the key to understanding the dust scale heights in this system. 
We model the polarization observations, focusing on the effects of dust scale heights. 
We find that the interplay between the inclination-induced polarization and the
polarization arising from radiation anisotropy in the crescent determines the observed polarization; the anisotropy is controlled by the dust optical depth along the midplane, which is, in turn, determined by the dust scale height in the vertical direction. 
We find that the dust grains can neither be completely settled nor well mixed with the gas. The completely settled case produces little radial displacement between the total and polarized flux, while the well-mixed case produces an azimuthal pattern in the outer (radial) edge of the crescent that is not observed. Our best model has a gas-to-dust scale height ratio of 2, and can reproduce both the radial displacement and the azimuthal displacement between the total and polarized flux. We infer an effective turbulence $\alpha$ parameter of approximately \todo{$0.0001-0.005$}. The scattering-induced polarization provides insight into a turbulent vortex with a moderate level of dust settling in the IRS 48 system, which is hard to achieve otherwise.
\end{abstract}

\section{Introduction}
Transition disks are a unique class of protoplanetary disks that have large inner dust cavities, typically tens of astronomical units (au) in size \citep{Espaillat2014,vanderMarel2023}. 
These cavities may indicate the presence of massive companions, such as planets, within the disk \citep{Artymowicz1996,Zhu2011,keppler2018,wagner2019}. 
In recent years, ALMA (Atacama Large Millimeter/submillimeter Array) observations have played a pivotal role in studying transition disks, providing high-resolution and sensitive imaging of their millimeter dust continuum emission. 
Previous works utilizing ALMA have revealed various intriguing features of transition disks, such as rings and gaps \citep{Fedele2017}, annular substructures \citep{Facchini2020},  and asymmetric structures \citep{vanderMarel2021_asymmetry}.
These observations have significantly contributed to our understanding of the structure, composition, and dynamics of transition disks, shedding light on the processes involved in their formation and evolution.


IRS 48 is a remarkable example among transition disks due to its unique very
prominent crescent-shaped structures \citep{vanderMarel2013}. 
One key aspect that sets IRS 48 apart is its well-studied grain properties. The multiwavelength observations of IRS 48 have revealed evidence of dust trapping, with a
significant difference in concentration among grains of different sizes \citep{vanderMarel2015}. 
Recently, deep high-resolution observations \citep{paper1} revealed an
eccentric dust ring that crosses the peak of the crescent-shaped structure.
This suggests an eccentric orbit of the dust grains, which is in line with
the asymmetric $^{13}$ CO (6-5) velocity maps \citep{vandermarel2016}. 
Both pieces of evidence suggest an (undetected) secondary companion in this system \citep{Calcino2019}.
Studying IRS 48 allows us to gain insight into the grain dynamics and trapping mechanisms in these disks, contributing to a more comprehensive understanding of their evolution.

Scattering-induced polarization at (sub)millimeter wavelengths has emerged as a powerful tool in the study of protoplanetary disks
\citep{stephens2014,stephens2017,Stephens2023,bacciotti2018,hull2018,cox2018,girart2018,harris2018,lee2018,Lee2021,
sadavoy2018,sadavoy2019,dent2019,harrison2019,aso2021,Ohashi2023,Tang2023}, offering valuable insights into various aspects of their composition and structure. 
This phenomenon, resulting from the self-scattering of dust grains, provides a unique avenue to explore crucial parameters such as dust grain sizes \citep{Kataoka2015,Kataoka2017,Yang2016a}, scale heights \citep{Yang2017,Ohashi2019}, and the composition or porosity of dust grains within these disks \citep{Tazaki2019,Yang2020,Zhang2023}. 
The observed self-scattering polarization patterns arise from at least two dominating mechanisms: inclination and anisotropy in disk structures. 
Inclination generates uniform polarization patterns along the minor axis of the disk, i.e. perpendicular to the position angle of the inclination.
This mechanism, first discussed by \cite{Yang2016a}, is responsible for most of the uniform polarization patterns observed so far. 
Anisotropy in disk structures, on the other hand, produces polarization perpendicular to disk substructures. This was first illustrated by \cite{Kataoka2015}, and the predicted polarization reversal
was later observed in HD142527 \citep{Kataoka2016}.

IRS 48 was previously observed with (sub)millimeter polarimetric observations by \cite{Ohashi2020}.
They observed uniform polarization patterns from self-scattering over roughly
a handful of beams. Together with nonpolarized multiband observations \citep{vanderMarel2015},
they put constraints on the sizes of dust grains and optical depth, favoring optically thick $\sim 100\rm\,
\mu m$ dust grains.
In this work, we present higher resolution polarimetric observations towards the IRS 48 system, aiming to understand the dust properties, especially the dust scale heights,
which are in turn related to the dynamics of the crescent-shaped structures and have far
reaching impact.

The paper is organized as follows. The observations are presented in Section~\ref{sec:results}. In Section~\ref{sec:model}, we model the system with Monte Carlo Radiative Transfer simulations, focusing on the effects of dust settling.
In Section~\ref{sec:discussion}, we discuss other implications of our results. 
The main results are summarized in Section~\ref{sec:summary}. 

\section{Observations and Results}
\label{sec:results}
\subsection{Observations}
\begin{figure*}[!htp]
\includegraphics[width=\textwidth]{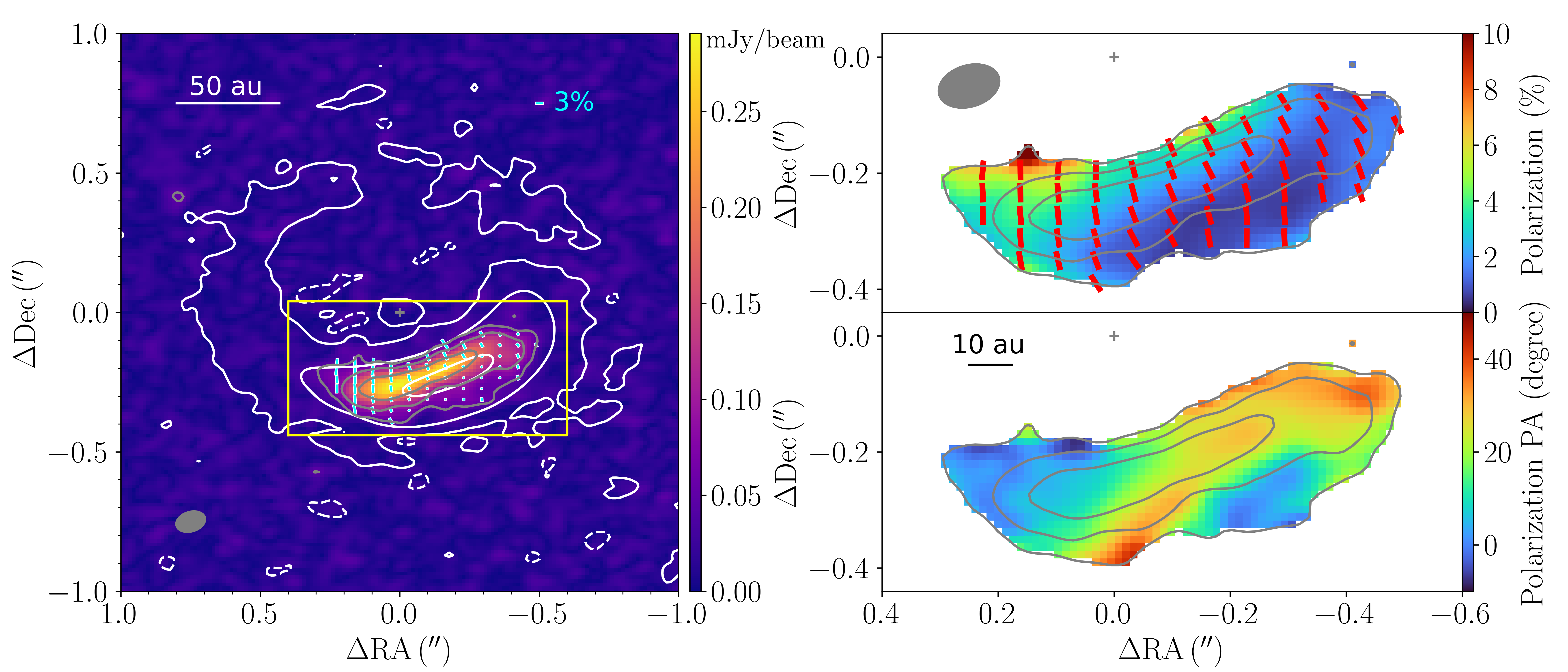}
\caption{Polarized dust continuum images. \textit{Left:} The colormap and the gray contours represent the polarized intensity in mJy/beam. Gray contour levels are plotted at the levels of $(4,8,16)\times \sigma_\mathrm{PI}$, with $\sigma_\mathrm{PI}=12\rm\, \mu Jy/beam$, which is the same across all three panels. The Stokes I image is shown with white contours plotted at the levels of $(-3, 3, 64, 1024)\times \sigma_\mathrm{I}$, with $\sigma_\mathrm{I}=14\rm\, \mu Jy/beam$. The cyan line segments represent polarization. The yellow rectangle with $1''\times 0\farcs48$ is the region of the right panels. The synthesized beam is $0\farcs11\times 0\farcs072$ with a position angle of $-73^\circ$. 
\textit{Upper right:} The polarization fraction, with red line segments of uniform length representing the polarization orientation.
Only regions with $>3\sigma$ detection are shown. 
\textit{Lower right:} The position angle of polarization orientation in degree. }
\label{fig:fig0}
\end{figure*}

The observations were carried out in June 7, June 14, and July 19, 2021, using ALMA Band 7 (0.87 mm) under the project code 2019.1.01059.S (PI: H. Yang). The details of the observation and the data reduction were presented in detail in \cite{paper1}. The final image is presented in Figure~\ref{fig:fig0}. The synthesized beam is $0\farcs{11}\times 0\farcs{072}$, which is $15\,\mathrm{au}\times 10\,\mathrm{au}$ (at a distance of 136 pc; \citealt{GaiaDR3}). The rms noise level measured in the Stokes I image is 14\,$\mu$Jy, whereas the rms noise level in Stokes\,QUV images is 12\,$\mu$Jy. The rms noise level of the (linearly) polarized intensity is $12\rm\, \mu Jy$.
 Stokes V is mostly below the detection limit, with only two dots in the east part of the crescent with
flux density peaks at $47\rm\, \mu Jy/beam$ and $42\rm\, \mu Jy/beam$, slightly larger than
$3\sigma$, but the areas of these two dots are much smaller than the beam size. 
We will ignore Stokes V throughout this paper and focus only on the linear polarization.

The inclination of the disk was constrained as $50^\circ$ with a position angle of $100^\circ$ \citep{Bruderer2014}.
To the zeroth order, the polarization orientation is broadly perpendicular to the position angle of the disk inclination,
$\mathrm{PA}\approx 10^\circ$ in the lower right panel of Figure~\ref{fig:fig0}. 
This is the expected polarization orientation of the inclination-induced self-scattering polarization \citep{Yang2016a}, 
which is also the mechanism adopted by \cite{Ohashi2020}. 

\begin{figure}
\includegraphics[width=0.5\textwidth]{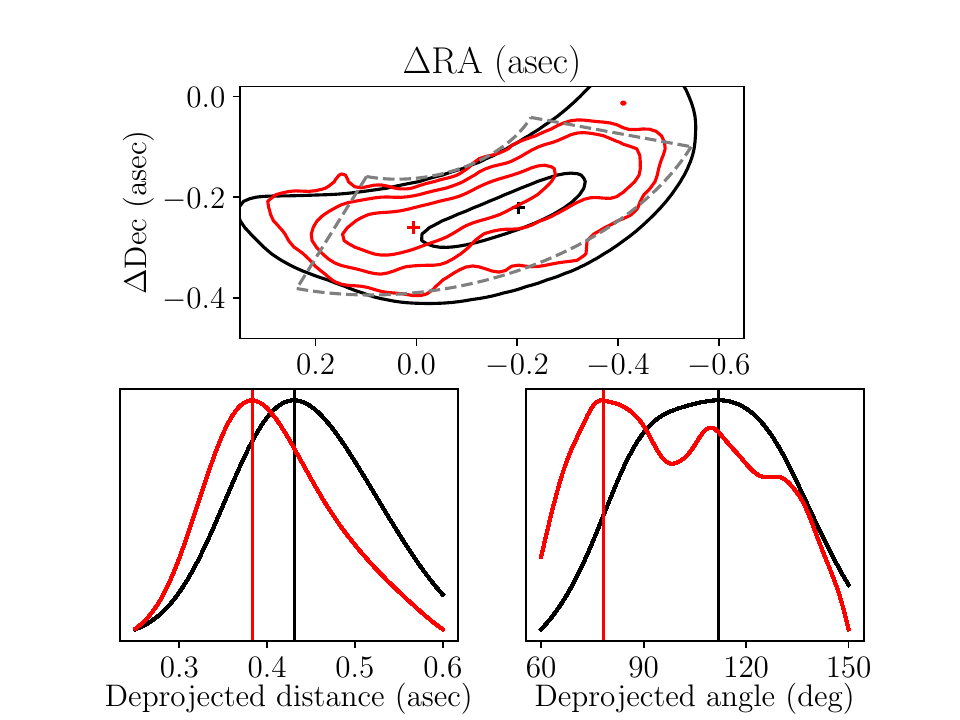}
\caption{\textit{Top}: The total flux (black contours) and the polarized flux (red contours).
The peaks for the total flux and the polarized flux are shown as red and black plus signs,
respectively. 
The grey fan-like region is enclosed by constant radius curves ($r=0\farcs 25$, $0\farcs6$) and constant $\phi$ lines ($\phi=60^\circ,$ $150^\circ$)
in the deprojected disk plane. 
\textit{Bottom Left}: Radial profiles (azimuthally averaged) of total flux (black) and polarized flux (red). Their peaks are labeled with vertical straight lines.
\textit{Bottom Right}: Azimuthal profiles (radially averaged) of total flux (black) and polarized flux (red). Their peaks are labeled with vertical straight lines.
Note that the total flux and polarized flux have different vertical scales.}
\label{fig:displacement}
\end{figure}

From the upper right panel of Figure~\ref{fig:fig0}, we can see that the polarization fraction is higher on the east side (left) and the inner radial part of the crescent.
This trend manifests itself as a displacement between the polarized intensity (PI),
shown as gray contours in the left panel, and the total intensity (I), shown as
white contours in the left panel. 
To make the displacement more evident, we plot the contours of the (polarized) intensity
as black (red) curves in the upper panel of Figure~\ref{fig:displacement}. 
We also plot a black (red) plus sign to mark the peak of the (polarized) intensity.
We can see that the peak of PI is displaced from that of I in both the radial and azimuthal
directions.
To quantify the difference, we take a wedge in the
(deprojected) disk coordinates defined with $0\farcs25-0\farcs6$ in the radial direction
and $60^\circ-150^\circ$ in the azimuthal angle in the disk frame, shown as a gray dashed fan in the upper panel of Figure~\ref{fig:displacement}.
We then integrate the fluxes along the azimuthal direction to obtain the radial profile,
shown in the lower left panel of Figure~\ref{fig:displacement}, and along the radial direction
to obtain the azimuthal profile, shown in the lower right panel. 
We can see that the peak of PI is inward of the peak of I by $0\farcs047$, or $6.4$
au. This displacement is significant, as it is much larger than the astrometric accuracy\footnote{According to ALMA Technical handbook (Cycle 9), the astrometric accuracy is about 9\% of the beam size. See their Section 10.5.2, and \cite{paper1} for more discussions.} 
of our observation, which is about $0\farcs012$ or $1.6$ au {and is deprojected as $0\farcs019$ or $2.5$ au in the disk plane}. The azimuthal displacement is
$34^\circ$.

\subsection{Significance of the observed radial and azimuthal displacements}
Before digging deep into the details and fitting the observations with models, we
would like to discuss qualitatively why the displacements are important first under the picture of
dust self-scattering. 
For an optically thick and geometrically thin layer of dust, the dust grains receive radiation only from other nearby dust grains. The polarization fraction saturates to a
constant value, independent of the spatial location. Indeed, \cite{Yang2017} calculated
the polarization fraction from a slab model, and the polarization fraction as a function of
optical depth reaches a plateau. If the dust in the IRS 48 crescent-shaped structure is
also geometrically thin (e.g., completely settled to the midplane), we expect roughly uniform polarization fraction, resulting in
coincidental peaks between PI and I. The observed displacements between the two therefore indicate that the dust responsible for the continuum emission in the crescent-shaped structure
may not be well settled towards the midplane. 
In the following sections, we will use radiative transfer modelings to quantitatively discuss the constraints on the dust distribution from the observed
displacements in detail. 

\section{Modeling}
\label{sec:model}
In this section, we model the polarization observations considering only the effects of self-scattering. 
We adopt a dust composition that is the same as the one adopted by \cite{Birnstiel2018}, which
has 0.2 water ice \citep{Warren2008}, 0.3291 astronomical silicate \citep{Draine2003}, 0.0743 troillite \citep{Henning1996}, and 0.3966 refractory organics \citep{Henning1996}, all in mass fractions.
At the wavelength of $870\rm\, \mu m$, this dust grain has a complex refractory index $m = n+ik$, with $n=2.3$ and $k=0.023$. 

\cite{paper1} and the left panel of Figure~\ref{fig:fig0} show that the dust grains are distributed along an
eccentric ring with a round head and a long tail { which are most visible in the lowest intensity contours}. However, most of the emissions come from the crescent-shaped
structure\footnote{The total flux contours are rather symmetric down to 64$\sigma$ (where most of the polarization detections are confined) in the deprojected view (see Figure 2 of \citealt{paper1})},
and polarized emission is detected only near the crescent. For simplicity, we assume a column density
of dust grains similar to that adopted by \cite{vanderMarel2015} and \cite{Ohashi2020}:
\begin{equation}
\Sigma_d(r,\phi) = \Sigma_0\exp\left(-\frac{(r-r_c)^4}{2r_w^4}-\frac{(\phi-\phi_c)^4}{2\phi_w^4}\right), 
\label{eq:Sigd}
\end{equation}
where $r_c=61\rm\, au$ and $\phi_c=109^\circ$ are radial and azimuthal locations of the peak, $r_w=10\rm\, au$
and $\phi_w=30^\circ$ are the radial and azimuthal extent, and $\Sigma_0$ is the column density at the peak.
Note that $\phi=0^\circ$ corresponds to the major axis of the projected disk, which has a position angle
of $100^\circ$ in the sky plane.

For the temperature structure of the disk, we adopt a simple power-law profile:
\begin{equation}
T_d = T_{60}\left(\frac{r}{60\rm\, K}\right)^{-0.5},
\end{equation}
where we take $T_{60}=60\rm\, K$ as the fiducial number, similar to the one adopted by \cite{Ohashi2020}, which was inspired by the physical-chemical model
developed by \cite{Bruderer2014}.

Our parameters are listed in Table.~\ref{tbl:pars}.
In this paper, we focus on the spatial distribution and profiles of the polarization fractions,
as opposed to the absolute polarization fractions. We fix $a_\mathrm{max}=140\rm\, \mu m$ to
represent grains of $\sim 100\rm\,\mu m$, because our main goal is to model the displacements, which are described by the profiles. The resulting polarization fractions in our models are of the same order as the observed ones. 
In addition to the aforementioned parameters, we define $f_\mathrm{settle}$ as the ratio 
between the scale heights of gas and dust. 
Based on its definition, $f_\mathrm{settle}=1$ implies a well-mixed grain model, whereas $f_\mathrm{settle}=\infty$ implies a perfectly settled dust layer.

\begin{table}
\centering
\begin{tabular}{|c|c||c|c|}
\hline
Parameter & Fiducial value & Parameter & Fiducial value \\
\hline
$r_c$ & 61 au & $r_w$ & 10 au \\
$\phi_c$ & $109^\circ$ & $\phi_w$ & $30^\circ$ \\
$\Sigma_0$ & $1.0\rm\, g/cm^2$ & $T_{60}$ & $60\rm\, K$ \\
$a_\mathrm{min}$ & $0.1\rm\, \mu m$ & $a_\mathrm{max}$ & $140\rm\, \mu m$ \\
$f_\mathrm{settle}$ & 2 & &\\ 
\hline
\end{tabular}
\caption{Parameters and their fiducial value in our model. See the text for their definitions.}
\label{tbl:pars}
\end{table}

{With these settings, we perform radiative transfer simulations using RADMC-3D\footnote{\url{https://www.ita.uni-heidelberg.de/~dullemond/software/radmc-3d/}}. 
The grid is spherical polar with 300, 128, and 128 points in the $r$, $\theta$,
and $\phi$ directions, respectively. The $r$-grids extend from 1 au to
300 au, evenly distributed in logarithmic space.
The $\theta$-grids extend between $\pi/2\pm 0.4$ uniformly.
The $\phi$-grids go around the entire $2\pi$ evenly. 
We then conduct Monte Carlo radiation transfer imaging with the assumed
temperature structure in full polarization mode with
$1.6\times 10^8$ photons, which is adequate according to the convergence test.
The results are convolved with the synthesized beam before being compared with the data.}

{In the following subsections, we investigate in detail the effects of the settling of dust grains. We first present our fiducial models and discuss the difference between the models, parameterized by $f_\mathrm{settle}$, and the data qualitatively in Section~\ref{ssec:th_settle}. This is followed by a quantitative analysis of the radial profiles in Section~\ref{ssec:th_rad} and the azimuthal profiles in Section~\ref{ssec:th_azi}. }

\subsection{Settling of dust grains}
\label{ssec:th_settle}

\begin{figure*}
\includegraphics[width=\textwidth]{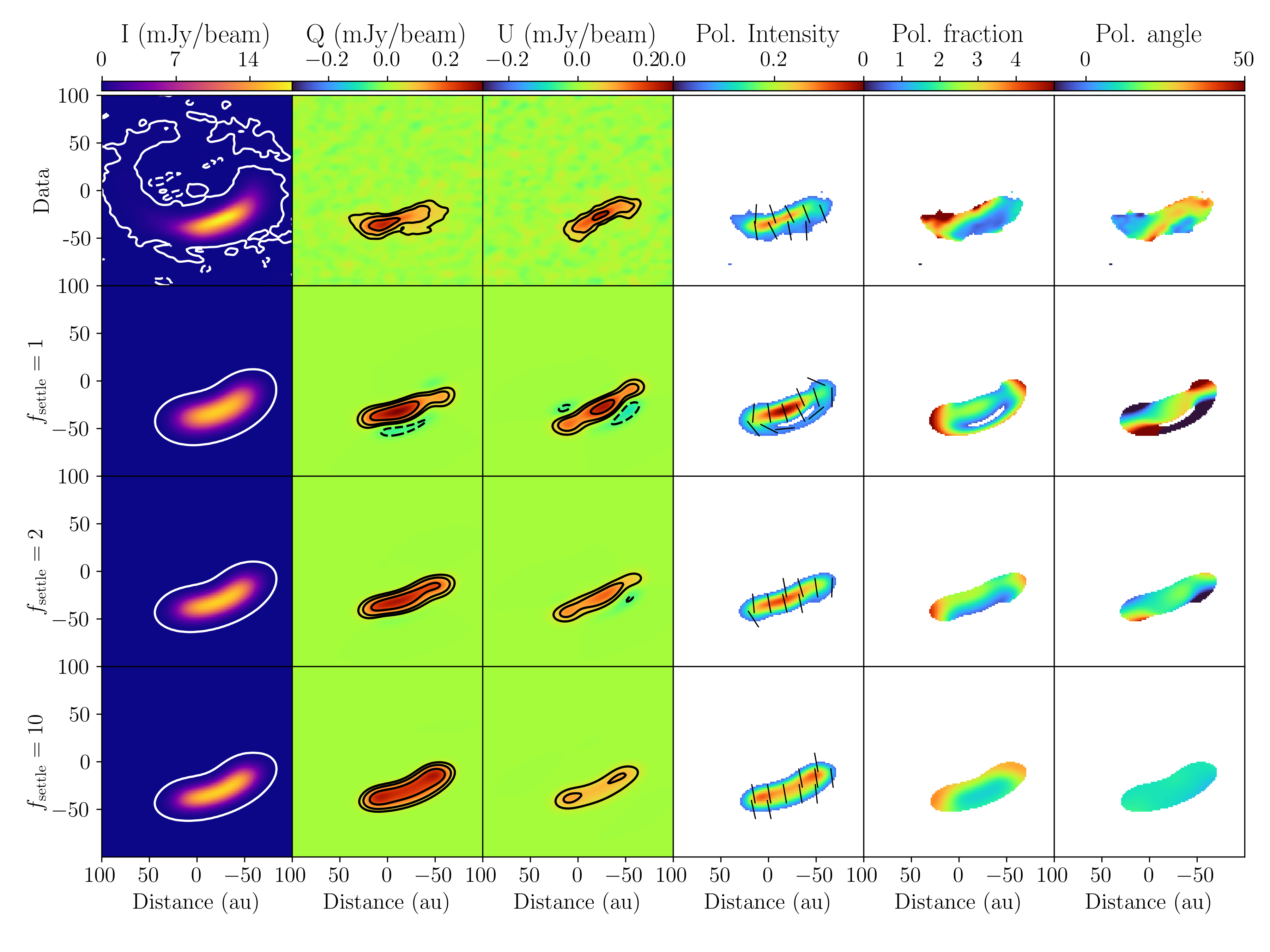}
\caption{Detailed comparisons between the data and our models. 
From top to bottom, each row represents the data and models with $f_\mathrm{settle}=1$, $2$, and $10$, respectively.
From left to right, each column represents Stokes I, Q, U, polarized intensity (PI), polarization fraction, and polarization position angle (PA), respectively. 
For the Stokes I column, the contours represent $\pm 3\sigma_\mathrm{I}$ contours. 
For the Stokes Q and U columns, the contours are plotted at the levels of $(-8, -4, 4, 8, 16)\times \sigma_{\rm PI}$. 
For the polarized intensity column, the line segments with uniform length represent the orientation of polarization.
For the right three columns, only regions with $PI>4\sigma_\mathrm{PI}$ are shown. 
}
\label{fig:combined}
\end{figure*}

We constructed a series of models with different values of $f_\mathrm{settle}$ and analyzed their corresponding polarized intensity, total intensity, polarization fraction, and angle. In Figure~\ref{fig:combined}, we present three models with $f_\mathrm{settle}=1,\, 2$, and $10$, together with the observation data.

Firstly, we would like to highlight that the total flux, Stokes I, exhibits minimal variation with respect to the parameter $f_\mathrm{settle}$, as demonstrated in the first column of Figure~\ref{fig:combined}. It should be noted that the $f_\mathrm{settle}=1$ model does display a slightly expanded structure with a greater radial extent, but this effect is relatively minor. On the contrary, the polarization map demonstrates a significantly higher sensitivity to changes in the $f_\mathrm{settle}$ parameter. 

For the model with $f_\mathrm{settle}=10$, i.e. the well-settled model, we observed that the polarized intensity and the total intensity exhibit similar distributions and that the polarization fraction and angle appear to be roughly uniform. 
The polarization fraction is slightly enhanced towards the two tips of the crescent.
This suggests that the polarization in this case is primarily influenced by the inclination-induced polarization.

On the other hand, for the model with $f_\mathrm{settle}=1$, i.e. the well-mixed model, we noticed a distinct feature in the polarization orientation. Specifically, the polarization shows azimuthal orientations in the (radial) outer edge of the crescent, which is not observed in our data. This phenomenon can be attributed to the fact that the dust in the disk puffs up, resulting in a smaller optical depth in the disk midplane. As a result, the dust particles in the outer region of the dust trap are illuminated by those in the inner region, leading to azimuthal polarization after scattering. However, this effect is not present in the model with $f_\mathrm{settle}=10$ due to the larger midplane optical depth, which prevents the light from penetrating from the inner to the outer region along the disk midplane.

These findings highlight the importance of considering the scale height ratio between gas and dust in understanding the polarization properties of astrophysical disks.

\subsection{Analysis of the radial profiles}
\label{ssec:th_rad}
\begin{figure}
\includegraphics[width=0.5\textwidth]{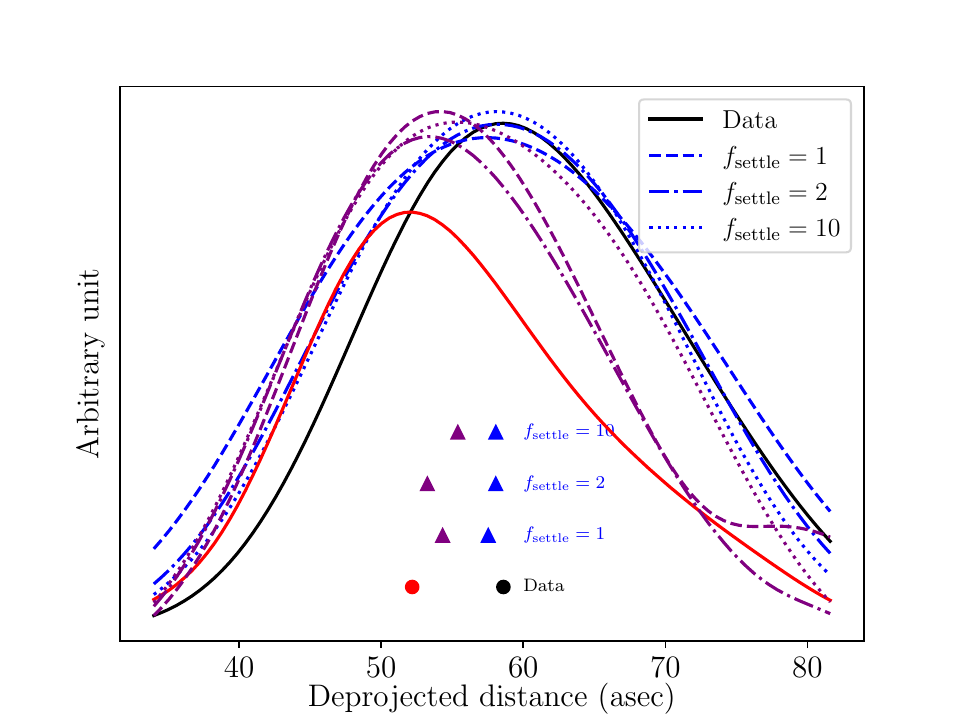}
\caption{The integrated radial profiles of (polarized) flux for the data and different models. The black (red) solid curves represent the total (polarized) flux of the data. The peak of the curve is marked as a black (red) dot. 
The blue (purple) curves represent the total (polarized) fluxes of the models. The peak of the curves
are marked as blue (purple) triangles. The $f_\mathrm{settle}=1$, 2, and 10 models are shown with 
dashed, dot dashed and dotted curves, respectively.}
\label{fig:rad-prof}
\end{figure}

To quantitatively study the effects of the parameter $f_\mathrm{settle}$, we performed flux integration to obtain the radial and azimuthal profiles. This procedure was carried out following the same methodology that was used to generate the observed total and polarized intensities shown in Figure~\ref{fig:displacement}. Figure~\ref{fig:rad-prof} displays the radial profiles for both total and polarized fluxes for the observed data, as well as for the three models. It is evident that all models exhibit broadly similar peaks in the deprojected radial distance. However, there is a noticeable difference in the peak positions between the total and polarized fluxes for each model. Specifically, the peak of the polarized flux is shifted inward (toward the star) compared to that of the total flux. In the case of the well-settled model with $f_\mathrm{settle}=10$, the shift is relatively small, measuring approximately $2.9$ au. 
On the contrary, for the less-settled models with $f_\mathrm{settle}=1$ and $2$, the radial shift is significantly larger, measuring approximately $3.3$ au and $4.7$ au, respectively.

The effects of $f_\mathrm{settle}$ on radial displacement can be explained by two factors. The first factor is the inclination-induced polarization, which generates a uniform polarization pattern. Due to the position angle and inclination of the system, the inclination-induced polarization produces a uniform polarization pattern with a position angle of $\sim10^\circ$. 
The second factor is the anisotropy of the radiation field in the 2D disk plane, which produces polarization that is roughly perpendicular to the intensity gradient. 
In the (radial) outer region, this polarization is in the azimuthal direction. The optical depth in the disk midplane plays a crucial role in determining which factor is more important. In the well-settled model with $f_\mathrm{settle}=10$, the optical depth is too large for light to travel along the disk midplane. As a result, all dust grains receive light only from nearby sources (i.e. neighboring dust grains), and the polarization pattern is solely determined by the inclination-induced polarization, resulting in a uniform polarization pattern and fractions. However, in the non-settled model with $f_\mathrm{settle}=1$, the radiation anisotropy in the 2D disk plane becomes important and produces complicated polarization patterns, as shown in the second row of Figure~\ref{fig:combined}.

The key idea is that the radiation anisotropy in the outer region of the disk leads to azimuthal polarization, which in turn reduces the polarization induced by inclination. This combination results in a shift of the peak of the polarized flux towards the inner region compared to the peak of the total flux. The magnitude of this shift is larger for less settled models.

When we increase the parameter $f_\mathrm{settle}$, which represents the settling of dust grains toward the disk midplane, the dust grains become more settled and the optical depth in the midplane increases. This increase in optical depth reduces the impact of radiation anisotropy on polarization. Consequently, the radial shift between the polarized flux and the total flux decreases.

\begin{figure}
\includegraphics[width=0.5\textwidth]{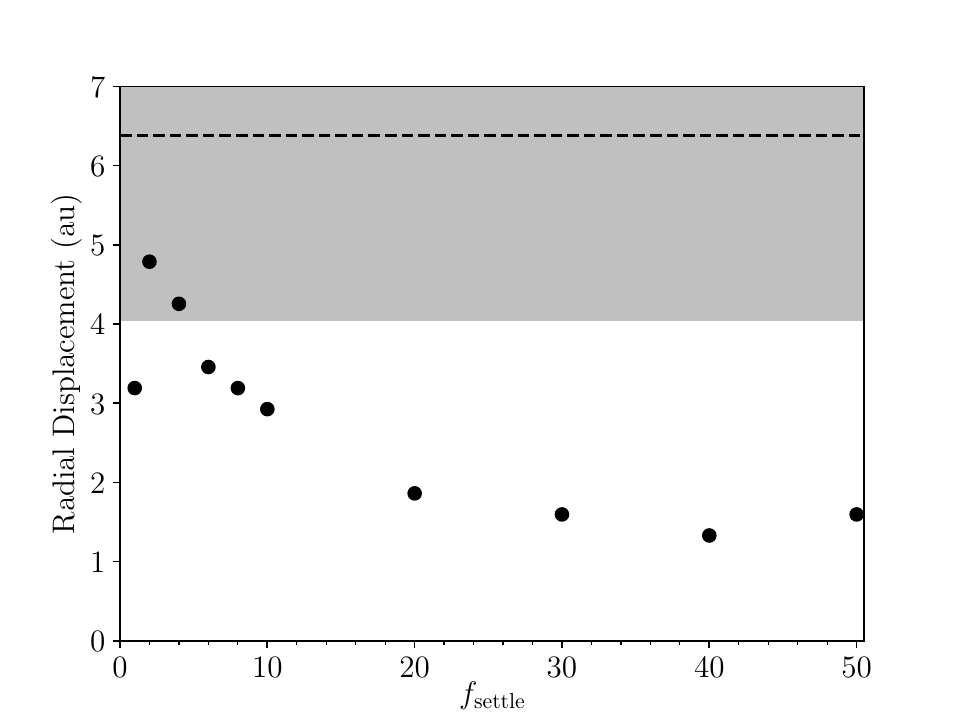}
\caption{The radial displacement (in the deprojected disk plane) between the peaks of the total flux and the polarized flux for models with varying $f_\mathrm{settle}$. The vertical dashed curve shows the radial displacement from our observation data ($6.4\rm\, au$). The gray region marks the \textbf{deprojected} astrometric accuracy around the observed value \todo{($6.4\pm 2.5\rm\, au$)}.}
\label{fig:dr_fset}
\end{figure}

In Figure~\ref{fig:dr_fset}, we have quantified this effect by plotting the radial displacement as a function of $f_\mathrm{settle}$. The dashed line represents the shift observed in our data, while the gray region indicates the astrometric accuracy. 
We can see that as we decrease $f_\mathrm{settle}$ from the well-settled case 
($f_\mathrm{settle}=10$), the displacement increases monotonically.
Decreasing $f_\mathrm{settle}$ makes midplane density smaller and the trend stops at $f_\mathrm{settle}=2$, beyond which the displacement starts to drop.
This is because the azimuthal polarization in the outer radial region starts to overwhelm the 
polarization from inclination, resulting in polarization patterns in the well-mixed model
($f_\mathrm{settle}=1$). The high polarization fraction, although in the ``wrong'' direction,
causes the PI peak to move radially outward, reducing the displacement between I and PI.
To reproduce the observed radial shift, we find that $f_\mathrm{settle}$ should be approximately $2$.

\subsection{Analysis of the azimuthal profiles}
\label{ssec:th_azi}
\begin{figure}
\includegraphics[width=0.5\textwidth]{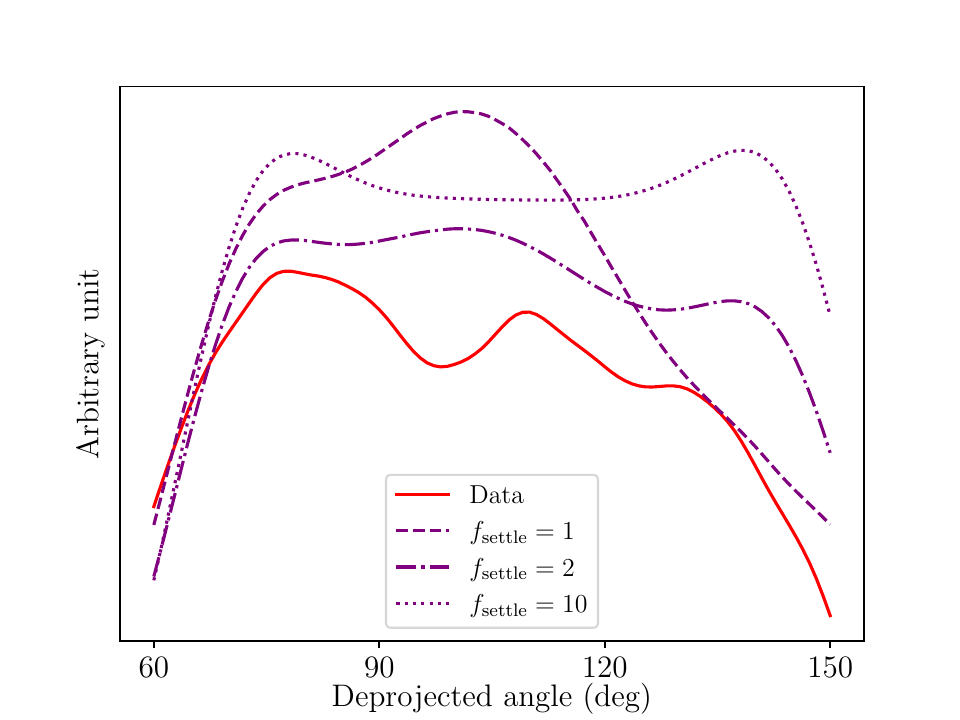}
\caption{The azimuthal profiles of polarized fluxes. The red solid curve represents the data. The purple curves represent the models. The azimuthal profiles of the total fluxes are similar and not shown in this figure. Note that the left and right sides are the same as the image
and represent east and west sides, respectively}
\label{fig:azi-prof}
\end{figure}

In Figure~\ref{fig:azi-prof}, we present the azimuthal profiles of the polarized flux for our observed data and three different models with $f_\mathrm{settle}$ values of 1, 2, and 10. The azimuthal profiles of the toal flux (Stokes I) exhibit remarkable similarity between the observed data and the three models, so we have not included them in the figure. For the model with $f_\mathrm{settle}=10$, the azimuthal profiles show two peaks near the outer edge. 
In particular, the profile demonstrates a rough symmetry between the eastern (to the left in Figure~\ref{fig:azi-prof}) and western regions. On the other hand, for the model with $f_\mathrm{settle}=1$, there is a prominent peak near the center of the crescent shape, slightly displaced by several degrees towards the east. In the case of the model with $f_\mathrm{settle}=2$, we observe three peaks, with the peak in the west being smaller than the one in the east. These two characteristics broadly resemble the characteristics observed in our data. 

The observed difference in polarization between the eastern and western parts of the moderately settled disk ($f_\mathrm{settle}=2$) can be explained by the interplay of two previously mentioned mechanisms. 
The first mechanism, inclination-induced polarization, generates a uniform polarization pattern with a PA of 10 degrees, perpendicular to the PA of 100 degrees of the major axis of the disk. This polarization pattern is consistent with the pattern in the $f_\mathrm{settle}=10$ model. The second mechanism, radiation anisotropy, produces polarization perpendicular to the intensity gradient. This results in a polarization pattern similar to the $f_\mathrm{settle}=1$ model in Figure~\ref{fig:combined}. In the eastern part, the PA of polarization ranges between 0 and 10 degrees, while in the western part, it is closer to 30-40 degrees. In the $f_\mathrm{settle}=2$ model, a combination of the $f_\mathrm{settle}=1$ and $f_\mathrm{settle}=10$ models is observed. In the eastern part, the two patterns have similar polarization PAs, which leads to their polarization adding together. However, in the western part, the two patterns have polarizations that make an angle of about 20-30 degrees, resulting in a significant cancelation. This cancellation is responsible for the azimuthal profiles observed in the $f_\mathrm{settle}=2$ model.

On the basis of our analysis, the $f_\mathrm{settle}=2$ model provides the best explanation for our observation. This model successfully generates a reasonable radial shift between the peaks of polarized flux and total flux. Additionally, it broadly accounts for the azimuthal profile of the polarization pattern, where the east side exhibits higher polarization compared to the west side. These characteristics arise from the interplay between inclination-induced polarization and radiation anisotropy. 

\section{Discussion}
\label{sec:discussion}
\subsection{Dust settling and level of turbulence}
The models in Section~\ref{sec:model} show that the crescent probably has a settling parameter
of $f_\mathrm{settle}=2$, i.e., the dust scale height is a factor of 2 smaller compared with
the gas scale height. This means that the dust grains are neither well mixed with the gas nor
completely settled towards the midplane. 

There are a couple of mechanisms that can be responsible for this moderate settling of dust grains, including
turbulent mixing \citep{Youdin2007}, pericenter oscillations in eccentric orbits \citep{Barker2014}, vertical flows in a vortex \citep{Lesur2009}, etc.
It is impossible to discuss
the true nature behind $f_\mathrm{settle}=2$, but we can get a quantitative sense of the
mixing, in terms of the turbulent parameter $\alpha$ \citep{Shakura1973}. In the most common scenario,
the gas-to-dust scale height ratio follows \citep{Youdin2007}:
\begin{equation}
    \frac{H_d}{H_g} = \left(1+\frac{\mathrm{St}}{\alpha}\right)^{-1/2}\left(\frac{1+2\mathrm{St}}{1+\mathrm{St}}\right)^{-1/2},
    \label{eq:Hd}
\end{equation}
where $\mathrm{St}$ is the Stokes number. For $f_\mathrm{settle}=2$, we have
\todo{$\alpha \approx  \mathrm{St}/3$}.
In the Stokes regime, we have $\mathrm{St}=\rho_s a /\Sigma_g$, where $\rho_s$ is the dust solid
density, $a$ is the grain size, and $\Sigma_g$ is the gas column density. For the adopted grain
size of $140\rm\, \mu m$, we have:
\begin{equation}
\alpha = 0.0014 \times \left(\frac{\rho_s}{3\rm\, g/cm^3}\right) \left(\frac{\Sigma_d}{1\rm\, g/cm^2}\right)^{-1} \left(\frac{\mathrm{G/D}}{10}\right)^{-1},
\end{equation}
{where $\Sigma_d=1\rm\, g/cm^2$ is the adopted dust column density, and G/D is the gas-to-dust ratio. }
The greatest uncertainty comes from the gas column density. The lower density of the gas column
requires a higher turbulence to stir the dust grains to a state with $f_\mathrm{settle}=2$.
{Nominally, the gas-to-dust ratio is $100$. In protoplanetary disks, especially transition disks with crescent structures, the G/D can be smaller than $10$ \citep{Ohashi2020}, or even as low as $3$ in the case of HD 142527 \citep{Soon2019}. For a G/D between $3$ and $100$, we have
the turbulence parameter as $\alpha = 0.0001\sim 0.005$.}

This estimate assumes compact spherical dust grains. 
It should be noted that porosity does not affect the inferred grain sizes too much.
In general, the peak wavelength of the scattering opacity remains the same as long as $f\times a_\mathrm{max}$ is fixed, where $f$ is the filling factor \citep{Tazaki2019}. So, the maximum
grain size that is responsible for the scattering-induced polarization roughly goes as
$1/f$. At the same time, the dust solid mass density is reduced to $f\rho_s$. These two effects
cancel each other out, which means that the inferred turbulent parameter $\alpha$ remains the same.

\subsection{Grain alignment}

The polarization pattern largely agrees with scattering-induced polarization, as shown in the models presented in Section~\ref{sec:model} and as discussed in \cite{Ohashi2020}.
Our models also reproduce the high polarization ($>3\%$) towards the eastern edge of the crescent. 

The observed polarization is inconsistent with the thermal emission by grains aligned with the existing mechanisms. 
The difficulties
come mostly from roughly uniform polarization patterns, which is hard to achieve
from mechanisms other than the inclination effects of self-scattering. 
Grains aligned with toroidal magnetic fields would produce polarization vectors that all
point towards the central star. If the magnetic fields possess a complicated eddy structure
or if the grains are aligned with eddy differential motions between gas and dust, we
would expect much more complicated polarization patterns. 
Radiative alignment, on the other hand, depends on the anisotropy in radiation flux
at the peak of the spectrum energy distribution inside the crescent-shaped structure. 
That is to say, the radiative alignment depends mainly on the light inside the crescent
at a wavelength of about $10$s of microns. 
Given that smaller grains are more azimuthally extended \citep{Geers2007, vanderMarel2013}, we would expect a
mostly radial radiation flux in the crescent, resulting in an azimuthal polarization
pattern. This is completely in contrast to our observation.

We can see that the theoretical expectations on grain alignment and their corresponding polarization patterns are very diverse and interesting. 
Our observation at $870\rm\, \mu m$ shows no signs of the aforementioned patterns 
from grain alignment.
Observations towards even longer wavelengths are needed to see whether and how grains
are aligned in this transition disk system. 
Even a spatially unresolved detection would be helpful. 
One beam of polarization patterns along radial (azimuthal) directions will rule out
radiative (toroidal magnetic) alignment. Complicated eddy morphology of magnetic fields or
aerodynamic alignment may result in nondetection at the crescent peak due to beam
averaging.

\section{Summary}
\label{sec:summary}

In this paper, we present high-resolution polarization observations towards the
transition disk IRS 48. The main findings are as follows.

\begin{itemize}
\item The polarization pattern is mostly uniform along a direction that is perpendicular
to the position angle of the major axis in the sky plane. 
This is in agreement with inclination-induced polarization.

\item The polarization fraction is mostly $1-2\%$. The east side is more polarized
than the west side, with a polarization fraction that reaches $3\%$ or higher. 

\item The peak of the polarized flux is displaced from that of the total flux, 
both radially inward and azimuthally eastward. 

\item We conduct radiative transfer calculations, focusing on the vertical settling of
dust grains. The results are shown in Figure~\ref{fig:combined}. 
We find that if the grains are well mixed with the gas, we would see azimuthal polarization
patterns at the outer part of the crescent. If the dust grains are completely settled, we would see
a uniform polarization pattern and very little radial displacement between the
total flux and the polarized flux. 

\item Our best model has a gas-to-dust scale height ratio of $2$. It can reproduce
the radial displacement and the azimuthal displacement simultaneously. 
This moderately settled model indicates that the crescent has an effective $\alpha$ of
about \todo{$0.0001\sim 0.005$}.

\item We did not find signs of grain alignment in our observation. 
Longer wavelength polarization observations may help determine whether the grains in the IRS 48 dust trap are aligned, and, if so, how. 
\end{itemize}

\section*{Acknowledgement}
{We thank the anonymous referee for their comments that help improve the manuscript. }
HY is supported by the National Key R\&D Program of China (No. 
2019YFA0405100) and the China Postdoctoral Science Foundation (No. 2022M710230).
ZYL is supported in part by NASA 80NSSC18K1095 and 80NSSC20K0533 and NSF AST-2307199.
LWL acknowledges support from NSF AST-1910364 and AST-2307844.
REH acknowledges support from NSF AST-1910364.

This paper makes use of the following ALMA data: ADS/JAO.ALMA\#2019.1.01059.S. 
ALMA is a partnership of ESO (representing its member states), NSF (USA) and NINS (Japan), 
together with NRC (Canada), MOST and ASIAA (Taiwan), and KASI (Republic of Korea), in  
cooperation with the Republic of Chile. The Joint ALMA Observatory is operated by  
ESO, AUI/NRAO and NAOJ.

\textit{Software:} CASA (v6.2.1.7; \citealt{casa}), RADMC-3D \citep{radmc3d}, Matplotlib \citep{Matplotlib}, Scipy \citep{scipy}

\bibliographystyle{aasjournal}

\end{CJK*}
\end{document}